\documentclass[sigconf]{acmart}
\AtBeginDocument{%
  }

\usepackage{multirow}
\usepackage{graphicx}
\usepackage{svg}
\usepackage{siunitx}
\usepackage{graphicx}
\usepackage{subcaption}

\copyrightyear{2026}
\acmYear{2026}
\setcopyright{othergov}
\acmConference[SIGIR '26]{Proceedings of the 49th International ACM SIGIR Conference on Research and Development in Information Retrieval}{July 20--24, 2026}{Melbourne, VIC, Australia}
\acmBooktitle{Proceedings of the 49th International ACM SIGIR Conference on Research and Development in Information Retrieval (SIGIR '26), July 20--24, 2026, Melbourne, VIC, Australia}
\acmDOI{10.1145/3805712.3809936}
\acmISBN{979-8-4007-2599-9/2026/07}

\begin{document}
% \raggedbottom

\title{GeoSearch: Augmenting Worldwide Geolocalization \\ with Web-Scale Reverse Image Search and Image Matching}

\renewcommand{\shorttitle}{GeoSearch: Augmenting Worldwide Geolocalization with Web-Scale Reverse Image Search and Image Matching}

\author{Tung-Duong Le-Duc}
\orcid{0009-0001-2603-0908}
\affiliation{
  \institution{University of Science, VNU-HCM}
  \city{Ho Chi Minh City}
  \country{Vietnam}
}
\email{23125081@student.hcmus.edu.vn}

\author{Hoang-Quoc Nguyen-Son}
\authornote{Corresponding authors.}
\orcid{0000-0003-2468-7815}
\affiliation{
  \institution{National Institute of Information \\ and Communications Technology}
  \city{Tokyo}
  \country{Japan}
}
\email{quoc-nguyen@nict.go.jp}

\author{Minh-Son Dao}
\authornotemark[1]
\orcid{0000-0003-3044-8175}
\affiliation{
  \institution{National Institute of Information \\ and Communications Technology}
  \city{Tokyo}
  \country{Japan}
}
\email{dao@nict.go.jp}

\renewcommand{\shortauthors}{Tung-Duong Le-Duc, Hoang-Quoc Nguyen-Son, and Minh-Son Dao}
%% No italics, no superscripts, not anonymous
%% Use footnote or author note to identify equal contribution and/or contact author info

\begin{abstract}
Worldwide image geolocalization, which aims to predict the GPS coordinates of any image on Earth, remains challenging due to global visual diversity. Recent generative approaches based on Retrieval-Augmented Generation (RAG) and Large Multimodal Models (LMMs) leverage candidates retrieved from fixed databases for reasoning, but often struggle with scenes that are absent from the reference set. In this work, we propose \textbf{GeoSearch}, an open-world geolocation framework that integrates web-scale reverse image search into the RAG pipeline. GeoSearch augments LMM prompts with database-retrieved coordinates and textual evidence extracted from web pages. To mitigate noise from irrelevant content, we introduce a two-layer filtering mechanism consisting of image matching, followed by confidence-based gating. Experiments on standard benchmarks Im2GPS3k and YFCC4k demonstrate the superiority of GeoSearch under leakage-aware evaluation. Our code\footnotemark[1] and data\footnotemark[2] are publicly available to support reproducibility.
\end{abstract}

\begin{CCSXML}
<ccs2012>
   <concept>
       <concept_id>10002951.10003317</concept_id>
       <concept_desc>Information systems~Information retrieval</concept_desc>
       <concept_significance>500</concept_significance>
       </concept>
 </ccs2012>
\end{CCSXML}

\begin{CCSXML}

\end{CCSXML}

\ccsdesc[500]{Information systems~Information retrieval}

\keywords{Image Geolocalization, Retrieval-Augmented Generation, Large Multimodal Models, Image Matching, Reverse Image Search}

\maketitle

% \vspace{-0.2cm}

\section{Introduction}
\label{sec:intro}

\footnotetext[1]{https://github.com/tungduong0708/GeoSearch}
\footnotetext[2]{https://huggingface.co/datasets/tduongvn/GeoSearch}

\begin{figure*}
  \centering
  \includegraphics[width=\textwidth]{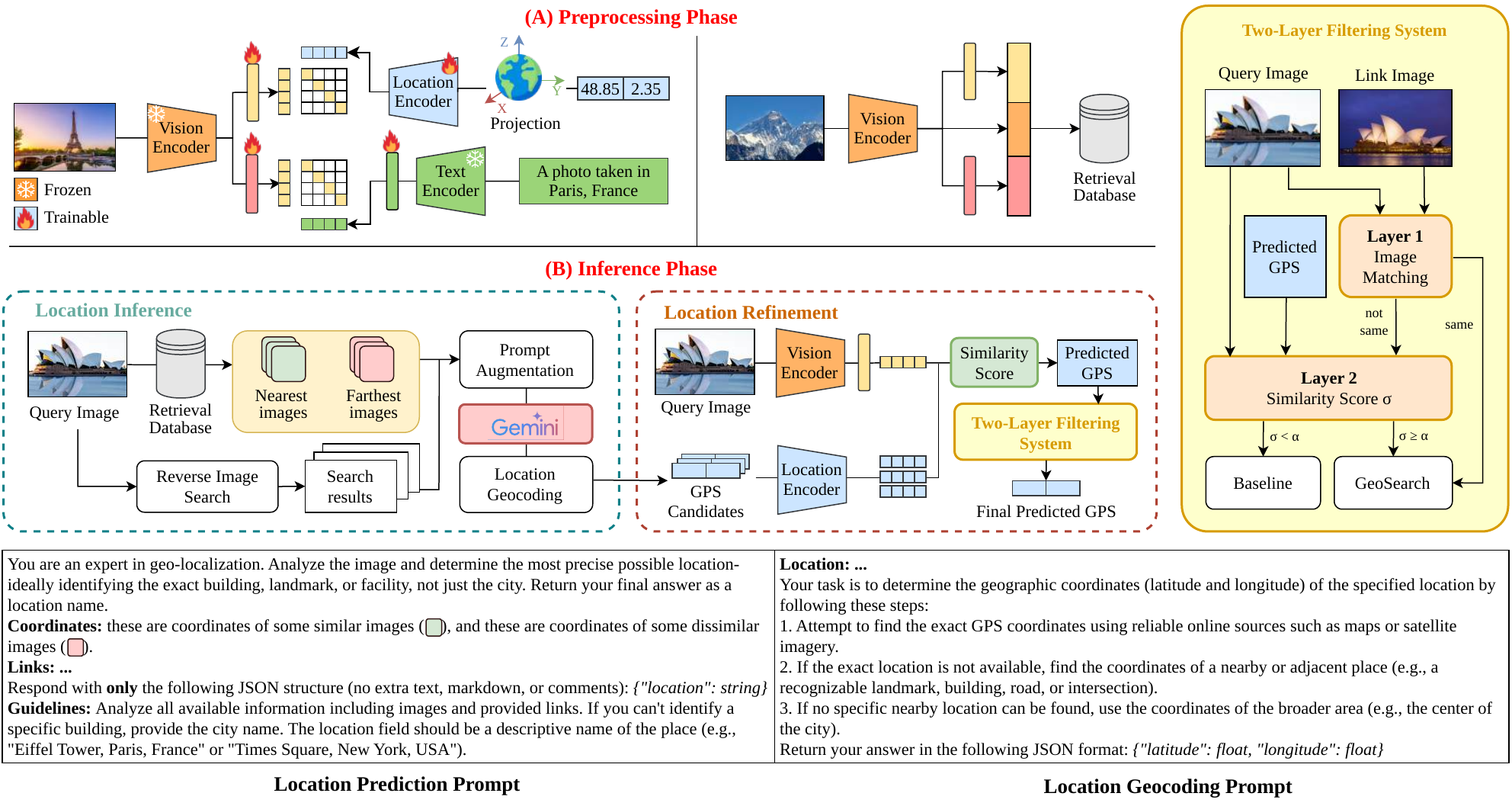}
  \vspace{-0.6cm}
  \caption{Overview of the GeoSearch framework.}
  \label{fig:overview}
  \vspace{-0.4cm}
\end{figure*}

Worldwide image geolocalization aims to predict geographic coordinates anywhere on Earth from visual content, enabling applications such as location-based services, navigation, and geographic information retrieval.
Recent generative methods based on Large Multimodal Models (LMMs) and Retrieval-Augmented Generation (RAG) \cite{zhou2024img2loc, jia2024g3, jia2025georanker} leverage retrieved context for reasoning, but remain limited by database coverage and retrieval quality, especially for newly constructed or unseen landmarks.

To address this limitation, we propose \textbf{GeoSearch}, a geolocation framework that integrates web-scale reverse image search into the RAG pipeline. GeoSearch builds upon G3~\cite{jia2024g3}, which learns unified image, text, and location representations for large-scale retrieval. We further adopt an Earth-Centered, Earth-Fixed (ECEF) projection to improve global geographic alignment. At inference time, GeoSearch augments LMM prompts with retrieved GPS coordinates and textual information from reverse image search results.

Since web-scale retrieval may introduce noisy or irrelevant results, GeoSearch employs a \textbf{two-layer filtering strategy} to adaptively select between search-augmented prediction and closed-world baseline prediction. First, image matching is performed using SuperPoint~\citep{detone2018superpoint} and LightGlue~\citep{lindenberger2023lightglue}. When matching fails, a confidence-based gating mechanism based on the similarity between the query image and the predicted GPS coordinates is applied, allowing the model to fall back to existing baseline methods.

Existing RAG-based approaches~\cite{zhou2024img2loc, jia2024g3, jia2025georanker} build databases from MP16~\cite{larson2017mp16} and evaluate on Im2GPS3k~\cite{hays2008im2gps} and YFCC4k~\cite{thomee2016yfcc100m}. As these datasets are derived from Flickr\footnotemark[3], this may cause data leakage. We therefore construct our database from OpenStreetView-5M (OSV-5M)~\cite{astruc2024openstreetview}, collected from Mapillary\footnotemark[4]. Under this leakage-aware setting, GeoSearch achieves state-of-the-art performance on both benchmarks and remains competitive even without relying on leaked data. This demonstrates the effectiveness of web-scale retrieval with adaptive filtering.

Our main contributions are summarized as follows:
\vspace{-0.08cm}
\begin{itemize}
    \item We propose an adaptive open-world geolocalization framework that integrates web-scale reverse image search with RAG-based LMM reasoning and two-layer filtering.
    \item We adopt the ECEF projection to improve global geographic alignment in location representation.
    \item We achieve state-of-the-art performance on both Im2GPS3k and YFCC4k under leakage-aware evaluation.
\end{itemize}

\footnotetext[3]{https://www.flickr.com/}
\footnotetext[4]{https://www.mapillary.com/}

\vspace{-0.2cm}

\section{Related Work}

\vspace{-0.08cm}

% \subsection{Worldwide Image Geolocalization}
\hspace*{0.8em} \textbf{Worldwide Image Geolocalization.}
Existing methods mainly follow three paradigms: classification, retrieval, and generative modeling. 
(1) \textit{Classification-based methods} \cite{muller2018isn, pramanick2022translocator, seo2018cplanet, clark2023geodecoder, haas2024pigeon, weyand2016planet, vo2017revisiting, izbicki2019exploiting} partition the Earth into discrete geographic regions and formulate localization as a multi-class prediction problem, but suffer from spatial quantization.
(2) \textit{Retrieval-based methods} match queries against geotagged images~\cite{khaliq2022multires, sferrazza2025match, zhu2022transgeo, shi2020looking} or GPS galleries~\cite{vivanco2023geoclip}. Visual retrieval, as adopted in visual place recognition (VPR), depends heavily on database coverage and diversity~\cite{sun2017dataset, warburg2020mapillary, torii201524, berton2022rethinking, barbarani2023local}, and degrades when similar locations are missing.
(3) \textit{Generative methods} include diffusion-based models~\cite{wang2025locdiffusion, dufour2025around} and LMM-based approaches~\cite{li2024georeasoner, li2025globe, wu2026geor, Shi_Li_Li_Fang_Zhou_Geng_Zhou_2026geobayes}. Recent RAG-based methods~\cite{zhou2024img2loc, jia2024g3, jia2025georanker} integrate retrieved references for localization, but operate in closed-world settings with static databases, limiting generalization to unseen landmarks.

% \subsection{Web-Scale Retrieval and Open-World Geolocation}
\textbf{Web-Scale Retrieval and Open-World Image Geolocation.}
Web-scale visual retrieval has been widely studied in landmark recognition and image indexing~\citep{wang2012towards, zheng2009tour} and is now deployed in reverse image search systems for Open-Source Intelligence and digital forensics~\citep{abdelnabi2022open,qi2024sniffer}. GeoSearch extends web-scale retrieval by incorporating automated reasoning for open-world geolocation.

\vspace{-0.2cm}
\section{Proposed Approach}

As shown in Figure~\ref{fig:overview}, GeoSearch has two phases: \textbf{(A) Preprocessing}, which learns unified image-text-location embeddings and builds a retrieval database; and \textbf{(B) Inference}, which combines database retrieval and reverse image search to generate GPS candidates, then applies similarity ranking and filtering for final prediction.

\vspace{-0.2cm}
\subsection{Preprocessing Phase}

\subsubsection{Model Architecture} 
Our model consists of three encoders for image, text, and location modalities.

\vspace{-0.1cm}
\paragraph{Image Encoder.}
We adopt the pre-trained CLIP \cite{radford20clip} ViT-L/14 as a shared visual backbone $V$. On top of this backbone, two parallel trainable transformation heads are applied to align image features with different modalities: given an input image $I$, the visual representation $V(I)$ is mapped to $\mathbf{e}^{img}_{txt} = f_{txt}(V(I))$ and $\mathbf{e}^{img}_{loc} = f_{loc}(V(I))$, where $f_{txt}$ and $f_{loc}$ are feed-forward projection layers for text and location alignment, respectively. 

\vspace{-0.1cm}
\paragraph{Text Encoder.}
For the text modality, a location description $D_{loc}$ is encoded by the CLIP text encoder $T$ and further projected by a trainable head $g_{txt}$ to obtain the text embedding $\mathbf{e}_{txt} = g_{txt}(T(D_{loc}))$.

\vspace{-0.1cm}
\paragraph{Location Encoder.}
To eliminate the geometric distortions inherent in 2D map projections used in prior work (e.g., Mercator~\cite{jia2024g3}, Equal Earth Projection (EEP)~\cite{vivanco2023geoclip}), we represent GPS coordinates using the \textbf{Earth-Centered, Earth-Fixed (ECEF)} system to preserve global spatial continuity. Given a coordinate with latitude $\phi$, longitude $\lambda$, and altitude $h$ (set to 0), we map it to a continuous 3D Cartesian vector $\mathbf{p} = [x, y, z]^T$:
\vspace{-0.05cm}
\begin{equation*}
\begin{aligned}
x &= (N(\phi) + h) \cos \phi \cos \lambda \\
y &= (N(\phi) + h) \cos \phi \sin \lambda \\
z &= ((1 - \varepsilon^2) N(\phi) + h) \sin \phi
\end{aligned}
\vspace{-0.15cm}
\end{equation*}
where $N(\phi)$ is the prime vertical radius of curvature and $\varepsilon$ is the Earth's eccentricity. 

To capture multi-scale geographic semantics, we adapt hierarchical Random Fourier Features~\cite{tancik2020fourier} to 3D ECEF coordinates. Specifically, given an ECEF vector $\mathbf{p}$, we apply a sinusoidal mapping $\gamma(\mathbf{p}, \sigma_i) = [\cos(2\pi \mathbf{W}\mathbf{p}), \sin(2\pi \mathbf{W}\mathbf{p})]^T$, where $\mathbf{W} \sim \mathcal{N}(0,\sigma_i)$ and $\sigma_i$ controls the frequency scale of the $i$-th layer. The resulting features are processed by scale-specific multi-layer perceptrons $h_i$ and aggregated across $K$ layers to produce the final location embedding
$\mathbf{e}_{loc} = \sum_{i=1}^{K} h_i(\gamma(\mathbf{p}, \sigma_i))$.

\vspace{-0.1cm}
\subsubsection{Training}

Following prior work~\cite{jia2024g3, vivanco2023geoclip}, we use a contrastive objective based on Information Noise-Contrastive Estimation (InfoNCE) to align image, text, and GPS representations. Let $a$ and $b$ denote two modalities with $\ell_2$-normalized embeddings $\mathbf{e}^a$ and $\mathbf{e}^b$. The loss from $a$ to $b$ is defined as
\vspace{-0.13cm}
\begin{equation}
\label{eq:infonce}
\mathcal{L}_{a,b}
=
-\frac{1}{|\mathcal{B}|}\sum_{i=1}^{|\mathcal{B}|}
\log
\frac{
\exp(\langle \mathbf{e}^a_i,\mathbf{e}^b_i\rangle/\beta)
}{
\sum_{j=1}^{|\mathcal{B}|}\exp(\langle \mathbf{e}^a_i,\mathbf{e}^b_j\rangle/\beta)
}
\vspace{-0.05cm}
\end{equation}

Here, $\langle\cdot,\cdot\rangle$ denotes cosine similarity, $\beta$ is the temperature set according to~\cite{jia2024g3}, and $|\mathcal{B}|$ is the batch size. We adopt a symmetric formulation and define the overall objective as
$
\mathcal{L}
=
\frac{1}{2}(
\mathcal{L}_{\mathrm{img},\mathrm{txt}}
+
\mathcal{L}_{\mathrm{txt},\mathrm{img}}
+
\mathcal{L}_{\mathrm{img},\mathrm{loc}}
+
\mathcal{L}_{\mathrm{loc},\mathrm{img}}
).
$

\vspace{-0.1cm}
\subsubsection{Database Construction}
The retrieval database is built by encoding each reference image into a unified multi-modal vector. For an image $I$, the stored representation concatenates raw visual features with text-aligned and location-aligned embeddings:
$
\mathbf{v}_{db} = \text{Concat}\left(
V(I),\;
\mathbf{e}^{img}_{txt},\;
\mathbf{e}^{img}_{loc}
\right).
$

\vspace{-0.2cm}
\subsection{Inference Phase}

\subsubsection{Location Inference}
Given a query image $I_q$, we first collect closed-world and open-world evidence. 

\begin{itemize}
    \item \textbf{Closed-World Retrieval.} Top-$k$ nearest and farthest neighbors of $I_q$ are retrieved from the database based on visual similarity, yielding a set of their GPS coordinates $\mathcal{G}_{db}$.
    
    \item \textbf{Open-World Retrieval.} We perform reverse image search via Google Lens and extract textual content from the retrieved web pages, denoted as $\mathcal{T}_{web}$.
\end{itemize}

\vspace{-0.2cm}
\paragraph{Prompt Construction and Generation.}

% Following G3~\cite{jia2024g3}, $n$ prompts ($n=4$) are constructed, each incorporating a different number of GPS coordinates, to obtain diverse location predictions.
% We further extend this design by integrating the top-$m$ web contexts from $\mathcal{T}_{web}$ ($m=5$) into each prompt and feeding them to \textbf{Gemini~2.0~Flash} to generate location descriptions. This design reduces quantization errors compared to direct coordinate generation.

Following G3~\cite{jia2024g3}, to obtain diverse GPS candidates for the Location Refinement stage, we construct $n$ prompts ($n=4$), each incorporating a different number of GPS coordinates to encourage prediction diversity. 
We further extend this design by incorporating content from the top-$m$ web contexts in $\mathcal{T}_{web}$ ($m=5$) into each prompt, providing richer contextual evidence.
The resulting prompts are fed into \textbf{Gemini~2.0~Flash} to generate location descriptions, which are later converted into GPS coordinates. 
This two-stage generation process reduces quantization errors compared to directly predicting coordinates.

\vspace{-0.1cm}
\paragraph{Location Geocoding.}
The generated locations are converted into GPS coordinates using OpenStreetMap Nominatim\footnotemark[5], with Gemini as a fallback. This forms a set of GPS candidates $\mathcal{C}_{gps}$.

\footnotetext[5]{https://nominatim.org/}

\vspace{-0.05cm}
\subsubsection{Location Refinement}
To determine the optimal candidate in $\mathcal{C}_{gps}$, each candidate $c \in \mathcal{C}_{gps}$ is embedded as $\mathbf{e}_{loc}(c)$ and matched with the query image $I_q$ using cosine similarity. The candidate with the highest score $\sigma$ is selected as the search-augmented prediction $P_{search}$ and passed to filtering.

\vspace{-0.05cm}
\subsubsection{Two-Layer Filtering System}
\label{filter}
To handle noisy web results, we employ a filtering mechanism to select between $P_{search}$ and the baseline prediction $P_{base}$ from existing methods, which relies only on closed-world database coordinates.

\vspace{-0.1cm}
\paragraph{Layer 1: Image Matching.}
We verify consistency between $I_q$ and the retrieved web image $I_{link}$ by extracting keypoints with \textbf{SuperPoint}~\cite{detone2018superpoint}, matching them using \textbf{LightGlue}~\cite{lindenberger2023lightglue}. A homography is then estimated using RANSAC, with a reprojection error threshold of $\tau_r$ pixels ($\tau_r = 4.0$). Let $M$ be the number of matched keypoints and $\rho$ be the inlier ratio after RANSAC. The image pair is considered to depict the same scene if $M \ge \tau_m$ and $\rho \ge \tau_{in}$ ($\tau_m = 50$ and $\tau_{in} = 0.5$). If satisfied, $P_{search}$ is accepted as final output; otherwise, the system proceeds to Layer~2. All filtering hyperparameters (e.g., $\tau_r$, $\tau_m$, $\tau_{in}$) are tuned on MP16-Search\footnotemark[6] (a validation set from MP16-Pro) via iterative evaluation and kept constant across experiments.

\footnotetext[6]{https://huggingface.co/datasets/tduongvn/MP16-Search}

\vspace{-0.1cm}
\paragraph{Layer 2: Confidence Gating.}
When matching fails or $I_{link}$ is unavailable, we rely on the model confidence by thresholding the similarity score $\sigma$ of $P_{search}$. If $\sigma \ge \alpha$ ($\alpha = 0.21$), $P_{search}$ is kept; otherwise, the system returns $P_{base}$. This guarantees that GeoSearch does not underperform the closed-world baseline in noisy scenarios.

\vspace{-0.15cm}
\section{Experiments}

\vspace{-0.05cm}
\subsection{Experimental Setup}

\paragraph{Dataset and Evaluation Metrics.}
As discussed in Section~\ref{sec:intro}, existing datasets such as MP16~\cite{larson2017mp16}, Im2GPS3k~\cite{hays2008im2gps}, and YFCC4k~\cite{thomee2016yfcc100m} are all derived from Flickr, raising concerns about data leakage. To address this issue, we build our retrieval database from the OSV-5M training set~\cite{astruc2024openstreetview}, which is sourced from Mapillary. For development, we split MP16-Pro~\cite{jia2024g3} into a 3,000-image validation set (\textbf{MP16-Search}) for hyperparameter tuning and a training set with the remaining images. Localization accuracy is reported based on geodesic distance at thresholds of 1, 25, 200, 750, and 2500 km.

\vspace{-0.1cm}
\paragraph{Training and Inference.}
We train the encoders on the MP16-Pro~\cite{jia2024g3} training split for 2 epochs using AdamW (learning rate $3\times10^{-5}$, weight decay $1\times10^{-6}$) with a batch size of 256. The temperature $\beta$ in Equation~\ref{eq:infonce} is set to 3.99. All experiments are conducted on a single NVIDIA H100 GPU. During inference, we use FAISS for nearest-neighbor retrieval and \textbf{Gemini Flash 2.0} for reasoning with temperature 1.0.

\vspace{-0.15cm}
\subsection{Comparison with SOTA Methods}

As shown in Table~\ref{tab:main_results}, we compare GeoSearch with recent RAG-based methods, including Img2Loc~\cite{zhou2024img2loc}, G3~\cite{jia2024g3}, and GeoRanker~\cite{jia2025georanker}, on Im2GPS3k and YFCC4k using the OSV-5M database. Under this leakage-aware setting, GeoSearch consistently achieves the best performance, particularly at fine-grained levels. On Im2GPS3k, GeoSearch with the GeoRanker baseline reaches 23.56\% at 1\,km and 89.59\% at 2500\,km, substantially surpassing all baselines. On YFCC4k, GeoSearch with the G3 baseline achieves the highest performance, reaching 17.53\% at 1\,km and 79.85\% at 2500\,km.

Furthermore, under the data leakage setting in Table~\ref{tab:leakage_im2gps3k}, although existing methods benefit from MP16-based retrieval, GeoSearch with the GeoRanker baseline still outperforms across all thresholds on Im2GPS3k, further validating its robustness.

\vspace{-0.2cm}
\begin{table}
\centering
\footnotesize
\begin{tabular}{l||S S S S S}
\hline
\textbf{Method}
& \multicolumn{1}{c}{\begin{tabular}[c]{@{}c@{}}\textbf{Street}\\\textbf{1 km}\end{tabular}}
& \multicolumn{1}{c}{\begin{tabular}[c]{@{}c@{}}\textbf{City}\\\textbf{25 km}\end{tabular}}
& \multicolumn{1}{c}{\begin{tabular}[c]{@{}c@{}}\textbf{Region}\\\textbf{200 km}\end{tabular}}
& \multicolumn{1}{c}{\begin{tabular}[c]{@{}c@{}}\textbf{Country}\\\textbf{750 km}\end{tabular}}
& \multicolumn{1}{c}{\begin{tabular}[c]{@{}c@{}}\textbf{Continent}\\\textbf{2500 km}\end{tabular}}
\\
\hline\hline

% === Im2GPS Section ===
\multicolumn{6}{c}{\textbf{Im2GPS3k} \cite{hays2008im2gps}} \\
\hline

{[L]}kNN, $\sigma=4$ \cite{vo2017revisiting} & 7.20 & 19.40 & 26.90 & 38.90 & 55.90 \\
PlaNet \cite{weyand2016planet} & 8.50 & 24.80 & 34.30 & 48.40 & 64.60 \\
CPlaNet \cite{seo2018cplanet} & 10.20 & 26.50 & 34.60 & 48.60 & 64.60 \\
ISNs \cite{muller2018isn} & 10.50 & 28.00 & 36.60 & 49.70 & 66.00 \\ 
Translocator \cite{pramanick2022translocator} & 11.80 & 31.10 & 46.70 & 58.90 & 80.10 \\ 
GeoDecoder \cite{clark2023geodecoder} & 12.80 & 33.50 & 45.90 & 61.00 & 76.10 \\
GeoCLIP \cite{vivanco2023geoclip} & 14.11 & 34.47 & 50.65 & 69.67 & 83.82 \\ 
PIGEON \cite{haas2024pigeon} & 11.30 & 36.70 & 53.80 & 72.40 & 85.30 \\
Img2Loc$^\dagger$ \cite{zhou2024img2loc} & 12.89 & 37.96 & 52.75 & 67.61 & 79.17 \\
G3$^\dagger$ \cite{jia2024g3} & 13.51 & 36.27 & 50.45 & 67.03 & 81.92 \\
GeoRanker$^\dagger$ \cite{jia2025georanker} & 10.08 & 36.50 & 54.79 & 72.87 & 86.65 \\
GLOBE \cite{li2025globe} & 9.84 & 40.18 & 56.19 & 71.45 & 82.38 \\
GeoBayes \cite{Shi_Li_Li_Fang_Zhou_Geng_Zhou_2026geobayes} & 6.30 & 34.70 & 53.60 & 73.70 & 85.90 \\ 
Geo-R \cite{wu2026geor} & 18.10 & 41.53 & 58.31 & 75.33 & 86.42 \\

\cline{1-6}

\textbf{GeoSearch (Ours)} & {} & {} & {} & {} & {} \\
\quad + Img2Loc$^\dagger$
& \textbf{23.56} & \underline{55.02} & \underline{66.97} & \underline{79.75} & 89.32 \\
\quad + G3$^\dagger$
& 23.49 & 54.92 & 66.87 & 79.58 & \underline{89.42} \\
\quad \textbf{+ GeoRanker}$^\dagger$
& \textbf{23.56} & \textbf{55.06} & \textbf{67.10} & \textbf{79.81} & \textbf{89.59} \\

\hline

% === YFCC Section ===
\multicolumn{6}{c}{\textbf{YFCC4k} \cite{thomee2016yfcc100m}} \\
\hline

{[L]}kNN, $\sigma=4$ \cite{vo2017revisiting} & 2.30 & 5.70 & 11.00 & 23.50 & 42.00 \\
PlaNet \cite{weyand2016planet} & 5.60 & 14.30 & 22.20 & 36.40 & 55.80 \\
CPlaNet \cite{seo2018cplanet} & 7.90 & 14.80 & 21.90 & 36.40 & 55.50 \\
ISNs \cite{muller2018isn} & 6.50 & 16.20 & 23.80 & 37.40 & 55.00 \\ 
Translocator \cite{pramanick2022translocator} & 8.40 & 18.60 & 27.00 & 41.10 & 60.40 \\ 
GeoDecoder \cite{clark2023geodecoder} & 10.30 & 24.40 & 33.90 & 50.00 & 68.70 \\ 
GeoCLIP \cite{vivanco2023geoclip} & 9.59 & 19.31 & 32.63 & 55.00 & 74.69 \\ 
PIGEON \cite{haas2024pigeon} & 10.40 & 23.70 & 40.60 & 62.20 & 77.70 \\
Img2Loc$^\dagger$ \cite{zhou2024img2loc} & 5.72 & 17.79 & 30.71 & 44.76 & 54.53 \\
G3$^\dagger$ \cite{jia2024g3} & 6.55 & 18.14 & 31.70 & 49.80 & 68.63 \\
GeoRanker$^\dagger$ \cite{jia2025georanker} & 5.20 & 18.21 & 33.95 & 55.49 & 73.41 \\
GeoBayes \cite{Shi_Li_Li_Fang_Zhou_Geng_Zhou_2026geobayes} & 4.90 & 16.10 & 30.90 & 55.80 & 75.40 \\ 
Geo-R \cite{wu2026geor} & 10.47 & 22.67 & 40.04 & 60.83 & 75.84 \\

\cline{1-6}

\textbf{GeoSearch (Ours)} & {} & {} & {} & {} & {} \\

\quad + Img2Loc$^\dagger$
& 17.48 & \textbf{35.21} & 48.15 & \underline{63.36} & 79.19 \\

\quad \textbf{+ G3}$^\dagger$
& \textbf{17.53} & \textbf{35.21} & \textbf{48.19} & \textbf{63.49} & \textbf{79.85} \\

\quad + GeoRanker$^\dagger$
& \underline{17.50} & 35.19 & \textbf{48.19} & \underline{63.36} & \underline{79.28} \\

\hline
\end{tabular}

% \caption{Main results on Im2GPS3k and YFCC4k. $^\dagger$ denotes models using OSV-5M for retrieval. Best results are in bold, and second-best are underlined.}
\caption{Main results on Im2GPS3k and YFCC4k. 
``+'' denotes integration of GeoSearch with baseline models at the filtering stage to select between GeoSearch and baseline output. 
$^\dagger$ denotes models using OSV-5M for retrieval. 
Best results are in bold, and second-best are underlined.}
\label{tab:main_results}
\vspace{-0.8cm}
\end{table}

\begin{table}
\centering
\footnotesize
\begin{tabular}{l||S S S S S}
\hline
\textbf{Method}
& \multicolumn{1}{c}{\begin{tabular}[c]{@{}c@{}}\textbf{Street}\\\textbf{1 km}\end{tabular}}
& \multicolumn{1}{c}{\begin{tabular}[c]{@{}c@{}}\textbf{City}\\\textbf{25 km}\end{tabular}}
& \multicolumn{1}{c}{\begin{tabular}[c]{@{}c@{}}\textbf{Region}\\\textbf{200 km}\end{tabular}}
& \multicolumn{1}{c}{\begin{tabular}[c]{@{}c@{}}\textbf{Country}\\\textbf{750 km}\end{tabular}}
& \multicolumn{1}{c}{\begin{tabular}[c]{@{}c@{}}\textbf{Continent}\\\textbf{2500 km}\end{tabular}}
\\
\hline\hline
Img2Loc    & 17.10 & \underline{45.14} & 57.87 & 72.91 & 84.68 \\
G3         & 16.65 & 40.94 & 55.56 & 71.24 & 84.68 \\
GeoRanker  & \underline{18.79} & 45.05 & \underline{61.49} & \underline{76.31} & \underline{89.29} \\
\textbf{GeoSearch}  & \textbf{23.56} & \textbf{55.06} & \textbf{67.10} & \textbf{79.81} & \textbf{89.59} \\
\hline
\end{tabular}
\caption{Results on Im2GPS3k under data leakage.}
\label{tab:leakage_im2gps3k}
\vspace{-0.8cm}
\end{table}

\vspace{-0.1cm}
% \vspace{-0.05cm}
\subsection{Ablation Study}

We conduct an ablation study on Im2GPS3k with GeoRanker baseline to analyze the effect of each component, as shown in Table~\ref{tab:ablation}. 

\textbf{w/o ClosedWorld:} This variant removes the closed-world retrieval component, relying only on reverse image search. Performance degrades across all distance thresholds, with larger drops at coarse levels. This highlights the importance of internal database candidates, which provide complementary geographic coverage when web retrieval alone is insufficient or noisy.

\textbf{w/o Location Geocoding:} Instead of generating a location description followed by geocoding, the LMM directly outputs GPS coordinates. While this slightly improves coarse-level accuracy, it causes a noticeable drop at street level (2.77\%). This suggests that the two-stage prediction process reduces coordinate noise and improves fine-grained localization.

\textbf{w/o Layer~1 (Image Matching):} In this variant, predicted GPS coordinates are passed directly to Layer~2. Similar to w/o Location Geocoding, this results in a slight decrease at fine-grained levels and a small increase at coarser levels. This indicates that Image Matching helps maintain candidate quality by filtering visually inconsistent matches early.

\textbf{w/o Layer~2 (Confidence Gating):} This variant disables the confidence-based filtering stage, causing the system to fall back to the baseline when candidates fail Layer~1. Its removal leads to substantial degradation across all distance thresholds, demonstrating its central role in filtering unreliable predictions and maintaining stable localization performance.

\vspace{-0.2cm}
\begin{table}
\centering
\footnotesize
\begin{tabular}{l||S S S S S}
\hline
\textbf{Method}
& \multicolumn{1}{c}{\begin{tabular}[c]{@{}c@{}}\textbf{Street}\\\textbf{1 km}\end{tabular}}
& \multicolumn{1}{c}{\begin{tabular}[c]{@{}c@{}}\textbf{City}\\\textbf{25 km}\end{tabular}}
& \multicolumn{1}{c}{\begin{tabular}[c]{@{}c@{}}\textbf{Region}\\\textbf{200 km}\end{tabular}}
& \multicolumn{1}{c}{\begin{tabular}[c]{@{}c@{}}\textbf{Country}\\\textbf{750 km}\end{tabular}}
& \multicolumn{1}{c}{\begin{tabular}[c]{@{}c@{}}\textbf{Continent}\\\textbf{2500 km}\end{tabular}}
\\
\hline\hline

GeoSearch
& \textbf{23.56} & \underline{55.06} & 67.10 & 79.81 & 89.59 \\
\hline
w/o ClosedWorld & 22.46 & 54.15 & 64.60 & 75.44 & 86.09 \\
w/o Geocoding
& 20.79 & 54.32 & \textbf{68.37} & \textbf{81.08} & \textbf{90.59} \\
w/o ImgMatch
& \underline{23.52} & \textbf{55.09} & \underline{67.13} & \underline{79.88} & \underline{89.69} \\
w/o ConfGate
& 19.85 & 48.25 & 62.60 & 76.38 & 88.25 \\
\hline
\end{tabular}
\caption{Ablation study on Im2GPS3k.}
\label{tab:ablation}
\vspace{-0.8cm}
\end{table}

\begin{table}
\centering
\footnotesize
\begin{tabular}{l||S S S S S}
\hline
\textbf{Method}
& \multicolumn{1}{c}{\begin{tabular}[c]{@{}c@{}}\textbf{Street}\\\textbf{1 km}\end{tabular}}
& \multicolumn{1}{c}{\begin{tabular}[c]{@{}c@{}}\textbf{City}\\\textbf{25 km}\end{tabular}}
& \multicolumn{1}{c}{\begin{tabular}[c]{@{}c@{}}\textbf{Region}\\\textbf{200 km}\end{tabular}}
& \multicolumn{1}{c}{\begin{tabular}[c]{@{}c@{}}\textbf{Country}\\\textbf{750 km}\end{tabular}}
& \multicolumn{1}{c}{\begin{tabular}[c]{@{}c@{}}\textbf{Continent}\\\textbf{2500 km}\end{tabular}}
\\
\hline\hline

\textbf{Img2Loc (MP16)}     
& \textbf{17.10} & \textbf{45.14} & \textbf{57.87} & \textbf{72.91} & \textbf{84.68} \\
Img2Loc (OSV-5M)   
& 12.89 & 37.96 & 52.75 & 67.61 & 79.17 \\
\hline
\textbf{G3 (MP16)}          
& \textbf{16.65} & \textbf{40.94} & \textbf{55.56} & \textbf{71.24} & \textbf{84.68} \\
G3 (OSV-5M)        
& 13.51 & 36.27 & 50.45 & 67.03 & 81.92 \\
\hline
\textbf{GeoRanker (MP16)}   
& \textbf{18.79} & \textbf{45.05} & \textbf{61.49} & \textbf{76.31} & \textbf{89.29} \\
GeoRanker (OSV-5M) 
& 10.08 & 36.50 & 54.79 & 72.87 & 86.65 \\
\hline
\textbf{GeoSearch (MP16)}   
& \textbf{23.92} & \textbf{55.49} & \textbf{67.77} & \textbf{80.28} & \textbf{90.19} \\
GeoSearch (OSV-5M) 
& 23.56 & 55.06 & 67.10 & 79.81 & 89.59 \\
\hline
\end{tabular}
\caption{Comparison of retrieval databases on Im2GPS3k.}
\label{tab:baseline_db_compare}
\vspace{-0.8cm}
\end{table}

\begin{table}
\centering
\footnotesize

\begin{tabular}{l||S S S S S}
\hline

\textbf{Method}
& \multicolumn{1}{c}{\begin{tabular}[c]{@{}c@{}}\textbf{Street}\\\textbf{1 km}\end{tabular}}
& \multicolumn{1}{c}{\begin{tabular}[c]{@{}c@{}}\textbf{City}\\\textbf{25 km}\end{tabular}}
& \multicolumn{1}{c}{\begin{tabular}[c]{@{}c@{}}\textbf{Region}\\\textbf{200 km}\end{tabular}}
& \multicolumn{1}{c}{\begin{tabular}[c]{@{}c@{}}\textbf{Country}\\\textbf{750 km}\end{tabular}}
& \multicolumn{1}{c}{\begin{tabular}[c]{@{}c@{}}\textbf{Continent}\\\textbf{2500 km}\end{tabular}}
\\
\hline\hline
GeoSearch (LLaVA)
& 17.08 & 42.08 & 54.05 & 67.23 & 79.01 \\
GeoSearch (GPT)
& \underline{21.72} & \underline{50.62} & \underline{61.26} & \underline{72.14} & \underline{83.98} \\
\textbf{GeoSearch (Gemini)}
& \textbf{23.56} & \textbf{55.06} & \textbf{67.10} & \textbf{79.81} & \textbf{89.59} \\
\hline
\end{tabular}
\caption{Impact of LMMs on GeoSearch (Im2GPS3k).}
\label{tab:lmm}
\vspace{-0.8cm}
\end{table}

\begin{figure}
    \centering
    \includegraphics[width=.9\linewidth]{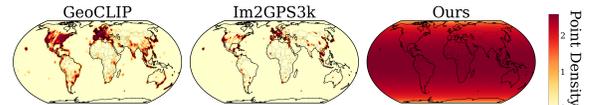}
    \vspace{-0.5cm}
    \caption{Geographic distributions of GPS galleries.}
    \label{fig:gps_gallery_dist}
\vspace{-0.5cm}
\end{figure}

\begin{table}
\centering
\footnotesize
\setlength{\tabcolsep}{4pt}
\begin{tabular}{l||S S S S S}
\hline
\textbf{Method}
& \multicolumn{1}{c}{\begin{tabular}[c]{@{}c@{}}\textbf{Street}\\\textbf{1 km}\end{tabular}}
& \multicolumn{1}{c}{\begin{tabular}[c]{@{}c@{}}\textbf{City}\\\textbf{25 km}\end{tabular}}
& \multicolumn{1}{c}{\begin{tabular}[c]{@{}c@{}}\textbf{Region}\\\textbf{200 km}\end{tabular}}
& \multicolumn{1}{c}{\begin{tabular}[c]{@{}c@{}}\textbf{Country}\\\textbf{750 km}\end{tabular}}
& \multicolumn{1}{c}{\begin{tabular}[c]{@{}c@{}}\textbf{Continent}\\\textbf{2500 km}\end{tabular}}
\\
\hline\hline
G3 Model
& \num{0.17} & \num{5.24} & 25.89 & 47.91 & 69.80 \\
GeoCLIP
& \textbf{0.20} & \underline{7.17} & \underline{29.10} & \underline{52.82} & \underline{74.11} \\
\textbf{GeoSearch Model}
& \textbf{0.20} & \textbf{9.34} & \textbf{31.13} & \textbf{56.89} & \textbf{74.94} \\
\hline
\end{tabular}
\caption{Performance on Im2GPS3k with GPS Gallery.}
\label{tab:gps_gallery}
\vspace{-0.85cm}
\end{table}

% \vspace{-0.05cm}
\vspace{-0.1cm}
\subsection{Effect of Retrieval Database}

Table~\ref{tab:baseline_db_compare} shows that most retrieval-based methods suffer noticeable performance drops on Im2GPS3k when using the OSV-5M database compared to MP16, revealing the impact of data leakage. Meanwhile, GeoSearch with the GeoRanker baseline remains competitive, with only a slight decrease under this leakage-aware setting.

\vspace{-0.2cm}
\subsection{Effect of LMMs}

We analyze the impact of different LMMs in GeoSearch with the GeoRanker baseline. As shown in Table~\ref{tab:lmm}, Gemini (Gemini Flash 2.0) performs best across all thresholds, followed by GPT (GPT-4o-mini) and LLaVA (LLaVA-Next-LLaMA3-8B).
% Despite a noticeable performance gap compared to stronger models, GeoSearch (LLaVA) remains competitive, particularly at fine-grained localization levels.

\vspace{-0.2cm}
\subsection{Effectiveness of ECEF Projection}

Following~\cite{vivanco2023geoclip}, we evaluate our ECEF-based encoder via image-to-GPS retrieval using cosine similarity, enabling comparison with GeoCLIP (EEP) and G3 (Mercator). Since GeoCLIP uses a 100K-point gallery with a distribution similar to Im2GPS3k, which may cause data leakage (Figure~\ref{fig:gps_gallery_dist}), we instead construct a uniformly distributed global gallery with 500K points for fair evaluation.

As shown in Table~\ref{tab:gps_gallery}, GeoSearch outperforms both baselines across all thresholds, with notable gains in coarse localization (+8.98\% at 750\,km), validating the ECEF projection's effectiveness.

% \vspace{-0.05cm}
\vspace{-0.2cm}
\subsection{Hyperparameter Analysis}

We analyze the hyperparameters of Layer~1 and Layer~2 under a unified binary classification between $P_{search}$ and $P_{base}$ on MP16-Search. Layer~1 thresholds are tuned on all queries and the failed cases are used to optimize Layer~2. Here, $P_{search}$ is obtained from the full pipeline with reverse image search, while $P_{base}$ relies on closed-world GPS candidates only. A case is labeled as \textit{Search Preferred} if $P_{search}$ outperforms $P_{base}$ in distance thresholds or geodesic error; otherwise, it is labeled as \textit{Baseline Preferred}.

\begin{figure}
  \centering

  % First row: Inlier ratio & Total matches
  \begin{minipage}{0.45\linewidth}
    \centering
    \includegraphics[width=\linewidth]{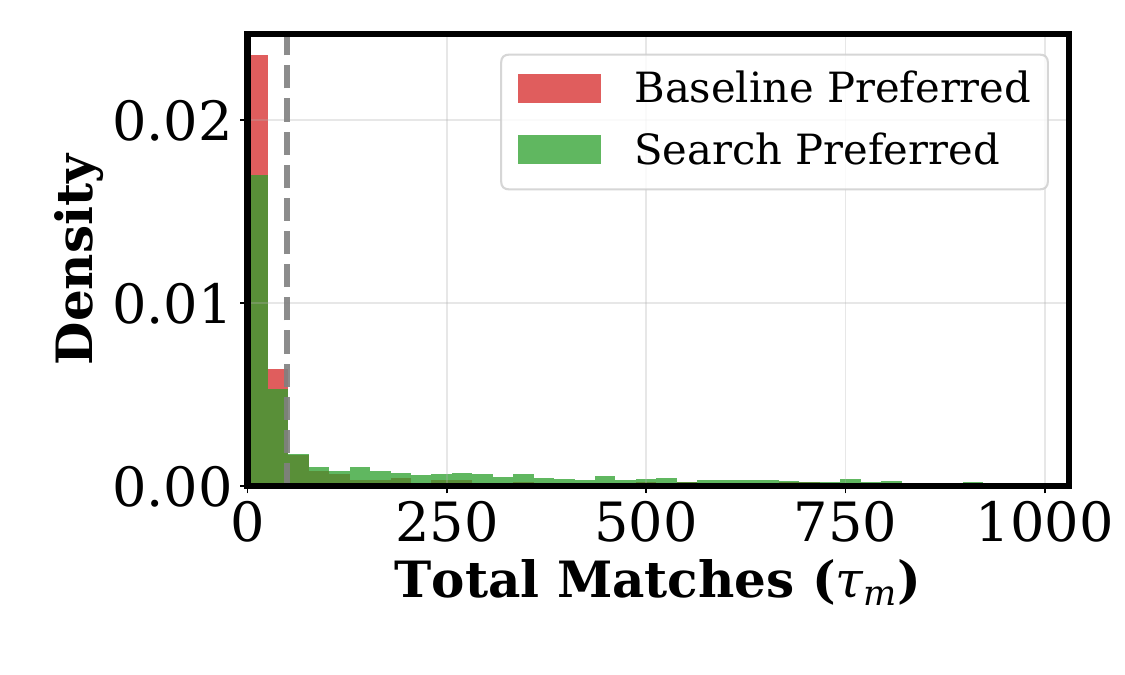}
  \end{minipage}
  \hfill
  \begin{minipage}{0.45\linewidth}
    \centering
    \includegraphics[width=\linewidth]{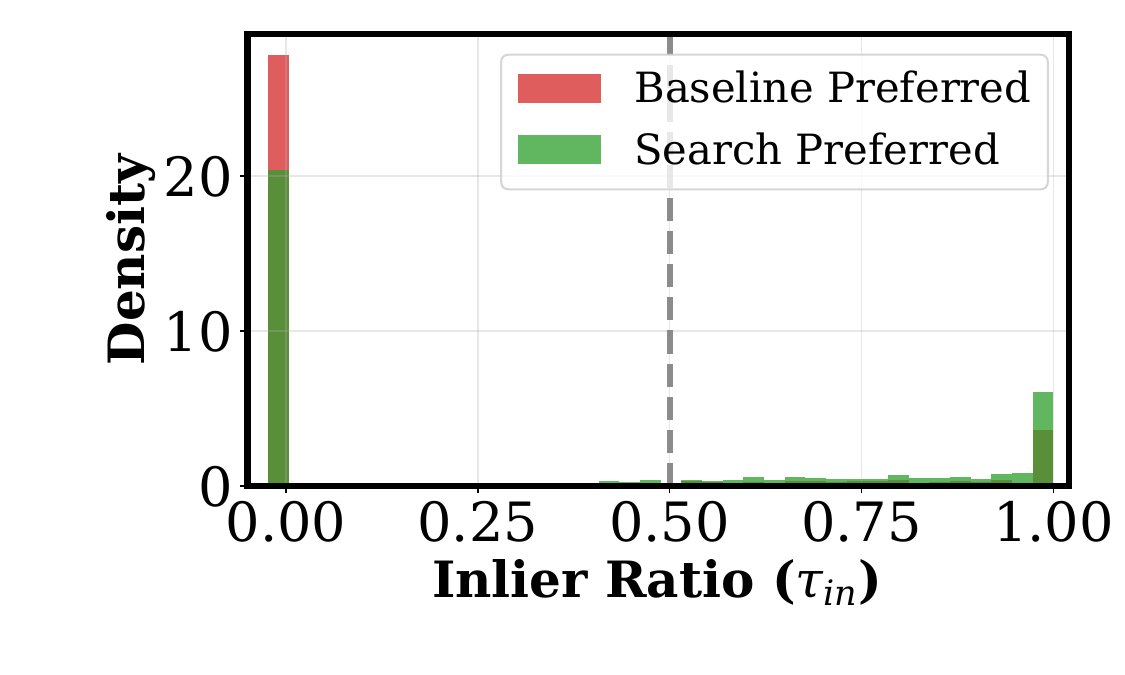}
  \end{minipage}

  % \vspace{0.1cm}

  % Second row: Layer2 Accuracy & F1
  \begin{minipage}{0.45\linewidth}
    \centering
    \includegraphics[width=\linewidth]{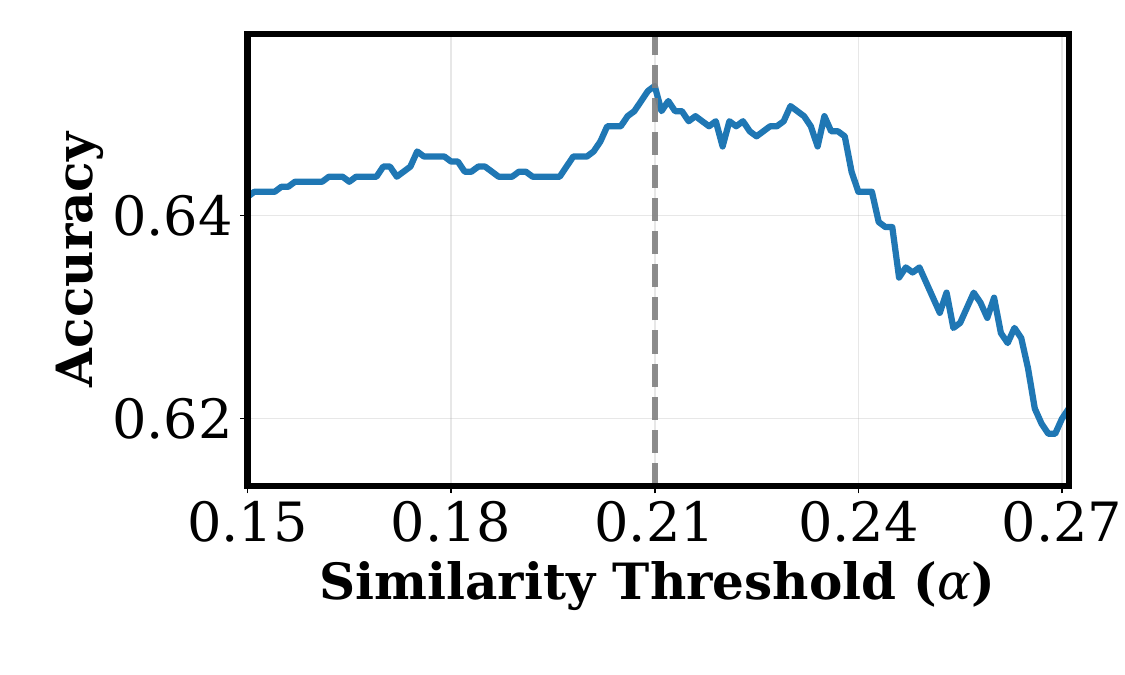}
  \end{minipage}
  \hfill
  \begin{minipage}{0.45\linewidth}
    \centering
    \includegraphics[width=\linewidth]{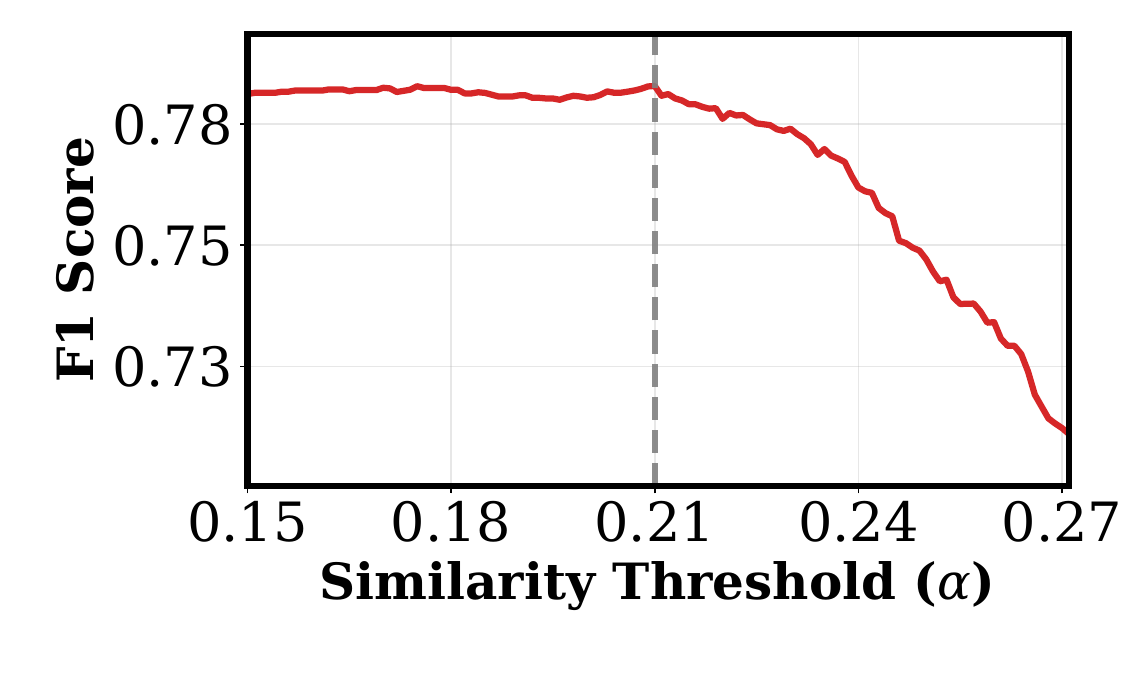}
  \end{minipage}

  \vspace{-0.4cm}
  \caption{Filtering hyperparameter analysis on MP16-Search.}
  \label{fig:threshold_tuning}
  \vspace{-0.1cm}
\end{figure}

As shown in Figure~\ref{fig:threshold_tuning}, $\tau_m=50$ and $\tau_{in}=0.5$ separate unreliable matches from reliable ones, where $P_{search}$ is more likely to outperform $P_{base}$. 
On the remaining queries after Layer~1, $\alpha=0.21$ yields the highest accuracy and F1-score over $\alpha \in [0,1]$, indicating reliable use of web evidence. These values are used in all experiments.

\begin{table}
\centering
\footnotesize
\setlength{\tabcolsep}{4pt}
\vspace{-0.2cm}
\begin{tabular}{l||S S S S}
\hline
\textbf{Method}
& \multicolumn{1}{c}{\textbf{Img2Loc}}
& \multicolumn{1}{c}{\textbf{G3}}
& \multicolumn{1}{c}{\textbf{GeoRanker}}
& \multicolumn{1}{c}{\textbf{GeoSearch}} \\
\hline\hline
Time (s)
& 2.48 & 2.70 & 4.48 & 5.96 \\
\hline
\end{tabular}
\caption{Average inference time (seconds).}
\label{tab:time}
\vspace{-0.9cm}
\end{table}

% \vspace{-0.05cm}
\vspace{-0.2cm}
\subsection{Efficiency Analysis}

\paragraph{Inference Time.}
As shown in Table~\ref{tab:time}, GeoSearch has higher latency than other RAG-based methods due to longer web-augmented prompts and the additional geocoding step for GPS extraction.

\vspace{-0.15cm}
\paragraph{Token Consumption.}
For location generation, the token usage is estimated as $462 + 19\,n_c + 2070\,n_l$, where 462 is the fixed prompt and image cost, and $n_c$ and $n_l$ denote the numbers of reference coordinates and retrieved web pages, respectively. Each web content is truncated to the first 2000 characters, and the average response is about 25 tokens. Fallback geocoding into GPS coordinates costs approximately 170 tokens with an average response of 35 tokens.

\vspace{-0.2cm}
\subsection{Failure Cases Analysis}

\begin{figure}[t]
  \centering
  \includegraphics[width=\columnwidth]{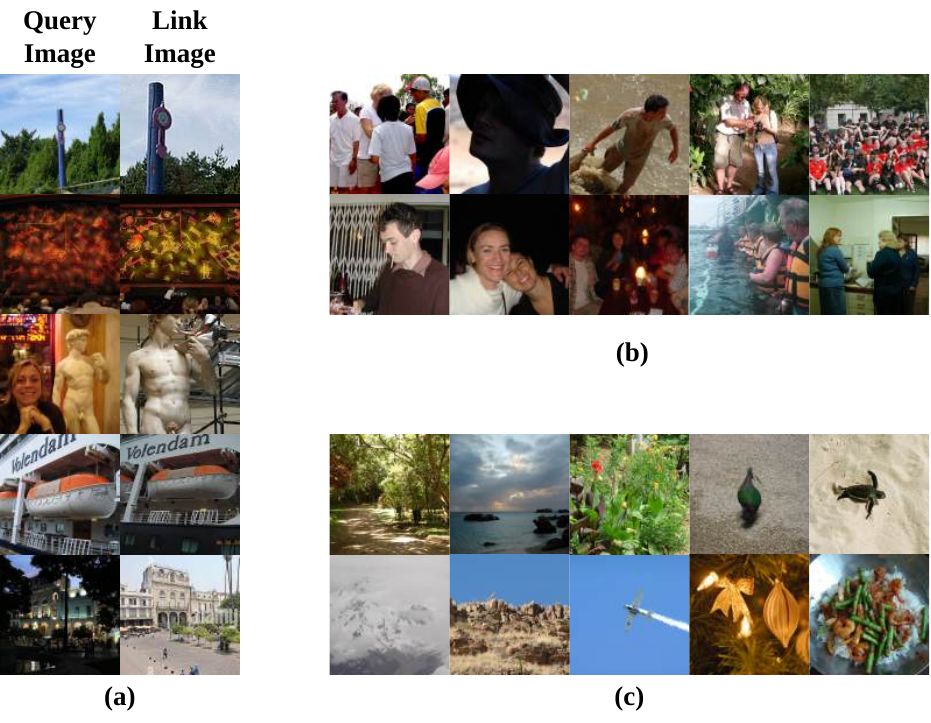}
  \vspace{-0.4cm}
  \caption{
  Failure cases in Im2GPS3k (geodesic distance $>$ 2500 km from ground truth). 
  (a) Query image with matched link image. 
  (b--c) Query images without reliable matches, including human-centered scenes or generic content (e.g., natural scenery or everyday objects).
  }
  \label{fig:failure}
  \vspace{-0.4cm}
\end{figure}

Despite strong performance, several challenging failure modes remain. 
Figure~\ref{fig:failure} presents representative examples from Im2GPS3k with large localization errors, grouped into two categories.

\textbf{Case 1: Matching link images exist but fail to localize correctly.}
As shown in Figure~\ref{fig:failure}(a), visually similar images can be retrieved from the Internet, but they do not provide sufficient structural or geographic cues to localize. Failures also occur when similar structures appear across multiple locations, leading to noisy and ambiguous web search results. Additionally, even with reliable matches, GeoSearch may still have low confidence in its prediction and fall back to the baseline output, resulting in incorrect predictions.

\textbf{Case 2: No reliable matching link images exist.}
In these cases, the query images lack distinctive location-specific content, making it difficult to retrieve useful matches from the Internet. 
Figure~\ref{fig:failure}(b) shows human-centered scenes dominated by ordinary people, where the background provides little contextual or geographic information. 
Figure~\ref{fig:failure}(c) presents natural or generic scenes, including vegetation, mountains, and miscellaneous objects such as food or vehicles, which are not specific enough to indicate a unique location. 
As a result, these images provide weak geographic signals and are difficult to localize using image retrieval alone.

% These failure modes highlight the importance of both reliable visual matches and strong location-specific cues, and suggest potential benefits from incorporating richer contextual signals such as semantic understanding or auxiliary metadata.

\vspace{-0.2cm}
\subsection{Limitations}

GeoSearch relies on web-scale image search for visual and textual context, which may raise privacy and ethical concerns. Moreover, extended prompts and external geocoding increase inference time compared to existing RAG-based methods.
Future work will explore more efficient prompt construction and geocoding, together with privacy-aware retrieval strategies, to mitigate these limitations.

\vspace{-0.2cm}
\section{Conclusion}

We proposed GeoSearch, an open-world image geolocalization framework that integrates web-scale reverse image search with retrieval-augmented multimodal reasoning. By leveraging external evidence, ECEF-based location encoding, and a two-layer filtering strategy, GeoSearch achieves state-of-the-art performance on Im2GPS3k and YFCC4k under leakage-aware evaluation. These results demonstrate its effectiveness and robustness on standard benchmarks. Further improvements will focus on efficiency and privacy while maintaining high accuracy at scale.

% \vspace{-0.2cm}
\begin{acks}
This research is supported by research funding from Faculty of Information Technology, University of Science, Vietnam National University - Ho Chi Minh City. 
\end{acks}

\newpage
\bibliographystyle{ACM-Reference-Format}
\bibliography{ref}

% \appendix

\end{document}